# Control of phase of the magnetization precession excited by circularly polarized fs-laser pulses

Alexander I. Chernov,[1,2] Mikhail A. Kozhaev,[1,2] Alexander N. Shaposhnikov,[4] Anatoly R. Prokopov,[4] Vladimir N. Berzhansky,[4] Anatoly K. Zvezdin,[1,2] Vladimir I. Belotelov[1,3,*]

[1] *Russian Quantum Center, 45, Skolkovskoye shosse, Moscow, 121353, Russia*
[2] *Prokhorov General Physics Institute RAS, 38 Vavilov Street, Moscow 119991, Russia*
[3] *Faculty of Physics, Lomonosov Moscow State University, Leninskie Gory, Moscow 119991, Russia*
[4] *Vernadsky Crimean Federal University, 4 Vernadskogo Prospekt, Simferopol, 295007, Russia*
*Corresponding author: belotelov@physics.msu.ru*

**Abstract** The inverse Faraday effect induced in magnetic films by ultrashort laser pulses allows excitation and control of spins at GHz and sub-THz frequencies. Frequency of the excited magnetization precession is easily tunable by the external magnetic field. On the other hand, phase of the precession hardly depends on magnetic field. Here we demonstrate an approach for the control of the precession phase by variation of the incidence angle of the laser pulses. In particular, theoretical consideration states that the phase increases with increase of the incidence angle and for small angles this relation is a direct proportionality. Experimental studies confirm this conclusion and provide shift of phase by about 4 deg. for the declination of the incidence angle by 15 deg. from the normal. This study provides a simple way for additional manipulation with optically excited magnetization dynamics, which is of importance for different spintronic applications.

## Introduction

All-optical excitation and detection of spin dynamics by ultrashort laser pulses have opened new ways for manipulation and control of spins at femto- and picosecond time scales [1–5]. In the case of the transparent magnetic films, laser pulses influence magnetization of the sample non-thermally by either inverse Faraday effect [6–8] or by the photoinduced magnetocrystalline anisotropy [7,9,10]. The former is observed for the circularly polarized pulses and can be interpreted in terms of the effective magnetic field induced within the sample during the pulse propagation. On the contrary, modifications of the magnetocrystalline anisotropy are observed for the linear polarized pulses.

The optical approach to the excitation of the magnetization dynamics overcomes several limitations inherent for the conventional methods utilizing microwaves. First of all, it allows to influence on the magnetization locally, within the focused laser spot, that can be easily shifted along the sample [11,12]. The micron size area of the sample where spins interact with photons can operate like a point source of magnons [13–15] . Variation of the laser spot shape and size provides tunability in terms of the type and spectrum of the generated spin waves. For example, one could switch between surface and backward volume spin waves by simply reducing diameter of the laser spot [16,17]. Moreover, passing from excitation with a single pulse to the multiple pulse excitation gives additional degree of freedom leading to enhancement of the spin waves amplitude, tunability of their spectrum and directivity of the magnon source [14,15].

Usually, optically excited magnetization dynamics is observed in the pump-probe experiment where high intensity pump beam drives spins and low intensity probe beam arrives at some time delay and measures variation of the magnetization along its wavevector by the Faraday effect. The magnetization precession at a given point is described by a decaying harmonic function: $m_z(t) = m_{z0} \sin(\omega t + \beta)$, where $m_{z0}$, $\omega$ and $\beta$ are precession amplitude, frequency and phase. The precession amplitude is easily increased by raising the excitation energy fluence while the frequency is changed via the external magnetic field. At the same time, control of the phase is not so straightforward. If magnetization dynamics is excited through the photoinduced magnetic anisotropy or inverse Cotton-Mouton effect then the phase can be modified by orientation of the linear polarization of the pumping beam [10,18,19]. However, in the case of the inverse Faraday effect dealing with circular polarized laser pulses only two phases of excited magnetization precession are available corresponding to left and right circular polarizations and they are 0 and $\pi$. Thus, tunable adjustment of the phase has not been demonstrated by now.

In this work we demonstrate the approach for the precise variation of the magnetization oscillation phase excited by circularly polarized femtosecond laser pulses. Theoretical investigation reveals that the phase depends on the light incidence angle and the relation is linear for small angles. The experimental studies confirm this behavior and are in good agreement with the theoretical model. The obtained results open a new way for the precise modification of the phase of the magnetization dynamics and, in particular, spin waves excited by circular polarized light pulses.

**Theory**

We start with the theoretical consideration of the magnetization dynamics excited within the illuminated spot of the film. Fig. 1 shows the considered system geometry, where the pump beam with the circular polarization enters the material in the ZY plane. It induces the effective magnetic field $\mathbf{H}_{IFE}$ directed along the **k** vector of light. The field $\mathbf{H}_{IFE}$ exists in the illuminated area of the magnetic film during the pulse propagation through the sample. Due to the pulse of $\mathbf{H}_{IFE}(t)$ magnetization of the sample becomes locally nudged and magnetization oscillations spread away in the form of spin waves. Therefore, generally, the magnetization dynamics should be described by the function $\mathbf{M}(\mathbf{r}, t)$. However, main properties of the optically induced magnetization dynamics within the illuminated spot can be considered in a simplified model which assumes a uniform precession of the magnetization $\mathbf{M}(t)$. In this model generation of spin waves can be taken into account by effective damping parameter $\alpha$ which exceed Gilbert parameter $\alpha_G$.

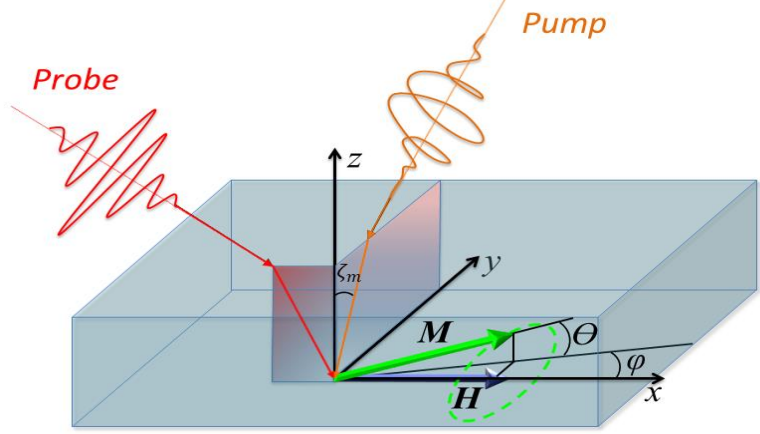

Fig. 1 System geometry. $\mathbf{H}$ is the external magnetic field. It also demonstrates the state of the system before the optical excitation. Circular polarized pump beam induces the magnetization dynamics resulting in precession that is depicted with green dashed line.

The magnetization dynamics is governed by the Landau-Lifshitz-Gilbert equation. In the spherical coordinate system with $z$-axis along the normal to the film and $x$-axis along the in-plane external magnetic field $H$ it is written by:

$$\frac{\partial \theta}{\partial t} + \alpha \sin\theta \frac{\partial \varphi}{\partial t} = -\frac{\gamma}{M}\frac{\partial U}{\partial \varphi}\frac{1}{\sin\theta}, \qquad (1a)$$

$$-\alpha \frac{\partial \theta}{\partial t} + \sin\theta \frac{\partial \varphi}{\partial t} = \frac{\gamma}{M}\frac{\partial U}{\partial \theta}, \qquad (1b)$$

where $\theta$ is the polar angle of magnetization, $\varphi$ is the azimuth angle of magnetization, $\gamma$ is the gyromagnetic ratio, $\alpha$ is the Gilbert damping constant and $U$ is the free energy density of the magnetic film. In the case of the predominant growth anisotropy with respect to the crystalline one the free energy density $U$ of the magnetic film is written by:

$$U = -(2\pi M^2 - K_U)\sin^2\theta - HM\sin\theta\cos\varphi - \mathbf{M}\mathbf{H}_{IFE}(t). \qquad (2)$$

The first term in Eq.(2) describes the magnetic anisotropy caused by the planar shape of the magnetic sample and its growth. The magnetic anisotropy can be characterized by the effective anisotropy field $H_a = 4\pi M - 2K_U/M$. If the in-plane magnetic field $H$ exceeds $H_a$ then in the absence of $\mathbf{H}_{IFE}$ the magnetization lies in-plane, which gives the equilibrium state $\theta_0 = \pi/2$ and

$\varphi_0 = 0$. Passing to $\theta_1 = \frac{\pi}{2} - \theta$, linearizing Eqs. (1) and taking Eq.(2) into account leads to the following set of equations:

$$-\frac{\partial \theta_1}{\partial t} + \alpha \frac{\partial \varphi}{\partial t} = -\omega_0 \varphi + \gamma H_{IFEy}(t) - \gamma H_{IFEx}(t)\varphi, \qquad (3a)$$

$$\alpha \frac{\partial \theta_1}{\partial t} + \frac{\partial \varphi}{\partial t} = -\omega_a \theta_1 - \omega_0 \theta_1 + \gamma H_{IFEz}(t) - \gamma H_{IFEx}(t)\theta_1, \qquad (3b)$$

where $\omega_a = \gamma H_a$ and $\omega_0 = \gamma H$. It follows from these equations that input of $H_{IFEx}$ in the excitation of the magnetization is negligibly small with respect to the input of the other $H_{IFE}$ components.

The problem of solving a set of non-homogeneous differential Eqs.(3) is equivalent to the problem of solving a set of homogeneous equations with initial conditions. Integration of Eqs. (3) over the duration of the pump pulse $\Delta t$ ($\Delta t \to 0$) gives the following initial conditions:

$$\theta_1(0+) = -\gamma H_{IFEy}\Delta t + \gamma \alpha H_{IFEz}\Delta t, \qquad (4a)$$

$$\varphi(0+) = \gamma H_{IFEz}\Delta t + \gamma \alpha H_{IFEy}\Delta t, \qquad (4b)$$

where $H_{IFEy,z}$ are amplitudes of the IFE field pulses: $H_{IFEy,z}(t) = H_{IFEy,z}\Delta t \delta(t)$ and it is assumed that $\alpha \ll 1$. The corresponding homogenous equation for $\theta_1$ is found from Eqs.(3) in the form:

$$\frac{\partial^2 \theta_1}{\partial t^2} + \frac{2}{\tau}\frac{\partial \theta_1}{\partial t} + \omega_r^2 \theta_1 = 0, \qquad (5)$$

where

$$\omega_r^2 = \omega_0(\omega_0 + \omega_a), \qquad (6)$$

$$\tau = 2/[\alpha(2\omega_0 + \omega_a)]. \qquad (7)$$

Solution of Eq. (5) has the form of

$$\theta_1 = \theta_m e^{-\frac{t}{\tau}} \sin(\omega_r t + \beta). \qquad (8)$$

Expressions for the amplitude $\theta_m$ and phase $\beta$ are obtained in view of Eqs.(4). Generally, they are rather cumbersome (Supplementary), but for $\alpha \to 0$ can be notably simplified:

$$\theta_m = \gamma \Delta t H_{IFEz}\sqrt{\frac{\omega_0^2}{\omega_r^2} + \tan^2 \zeta_m}, \qquad (9)$$

$$\tan \beta = \frac{\omega_r}{\omega_0} \tan \zeta_m, \qquad (10)$$

where we take into account that $\frac{H_{IFEy}}{H_{IFEz}} = \tan \zeta_m$, $\zeta_m$ is refraction angle of light inside the magnetic film. Therefore one can see that phase of the magnetization precession increases when angle of light incidence, $\zeta_i$ ($\sin \zeta_i = n \sin \zeta_m$, $n$ is refraction index of the magnetic film), gets larger. If $\zeta_m$ is relatively small ($\zeta_m \ll 1$) then $\beta$ becomes linearly dependant on $\zeta_m$. For the magnetic films with in-plane uniaxial magnetic anisotropy and in relatively small magnetic fields

$\omega_r/\omega_0$ might be rather large that provides notable variation of the phase even for moderate incidence angles.

**Experimental**

For the experimental investigation we have used the rear-earth iron-garnet film with bismuth ion substitution ($Bi_{1.4}Y_{1.6}Al_{1.55}Sc_{0.2}Fe_{3.25}$). It was grown by liquid phase epitaxy on the gadolinium gallium garnet substrate with a crystallographic orientation (111). The film thickness is 4.1 μm, the saturation magnetization is $4\pi M_s$ = 240 G, the uniaxial anisotropy constant $K_u$ is negative and equals to -10 erg/cm$^3$, while γ is 1.7·10$^7$ s$^{-1}$ Oe$^{-1}$.

The sample was studied by the two color pump-probe technique. The pump wavelength is 616 nm, pump pulses are circularly polarized and cause the magnetization precession. The light energy fluence is 0.2 mJ/cm$^2$ (calculated for 9 μm beam diameter). The probe pulse wavelength is 820 nm and has 15 times less energy fluence. Both pump and probe 150-fs pulses are generated by Newport Mai Tai HP Ti:Sapphire laser and Spectra-Physics Inspire Auto 100 optical parametric oscillator at 80.54 MHz repetition rate. Faraday effect for the transmitted probe pulses is used for the detection of the magnetization precession (Nirvana balanced diode detector with lock-in amplifier). The time delay between the pump and probe pulses is varied from -0.2 to 2.5 ns, where zero-time delay corresponds to the simultaneous propagation of the pump and probe pulses through the sample.

Focusing of the pump and probe beams was performed by the single reflective microscope objective. The incidence angle of the probe beam was set constant to 17 deg. in the orthogonal plane XZ. While for the pump beam we have used angles between -15 to 15 deg. and the orthogonal plane YZ (Fig. 2a). The change of the incidence angle for the pump beam was achieved by moving the entrance position of the beam within the objective and using various sectors of the objective.

**Results and Discussion**

The magnetization precession excited by laser pulses in different external magnetic fields from 15 Oe to 465 Oe is shown in Fig. 2b. Its temporal dependence is well described with the decaying harmonic function by Eq.(8). The decay time of the magnetization precession is 13.5 ns, which in accordance to Eq.(7) corresponds to the Gilbert damping parameter $\alpha$ = 0.002. The oscillation frequency grows almost linearly with magnetic field (inset in Fig. 2b). It is also in agreement with theory, in particular with Eq.(6). Some deviation from Eq.(6) appears for realtively small magnetic fields $H$ < 20 Oe which can be accounted for excitation of the spin waves.

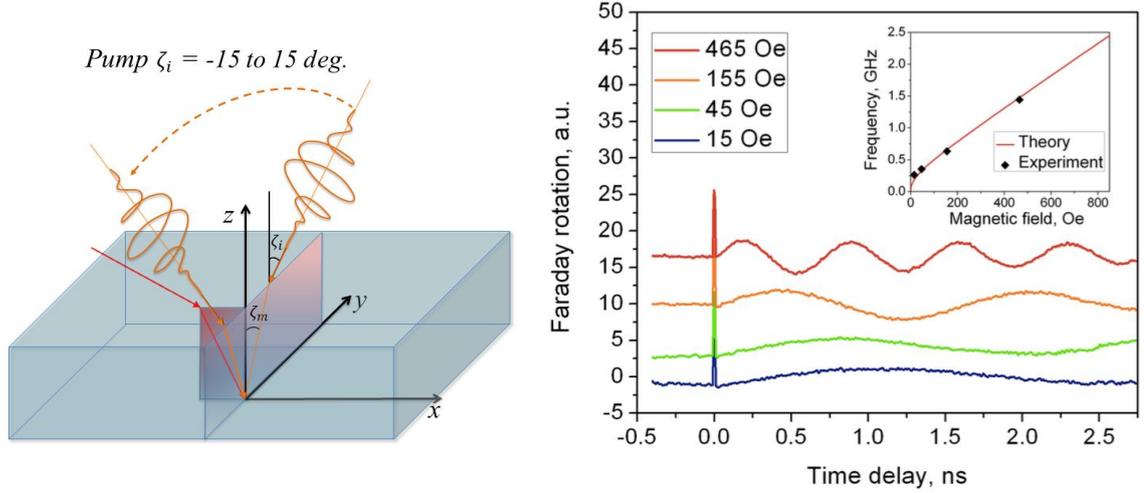

Fig. 2. a. Scheme of the experiment. The pump beam always hits the sample in the YZ plane. b. Normalized time-resolved change of the Faraday rotation indicating the magnetization precession excited by laser pulses in different external magnetic fields. Inset: dependence of the precession frequency on the external magnetic field.

Let us now vary the incidence angle of the pump and pay attention to the phase of the excited magnetization precession (Fig. 3a). One can notice the trend of the phase change with variation of the incidence angle. If the observed magnetization precession curves are fitted by Eq.(8) then the phase dependence on the incidence anlge is quantified (red circles in Fig. 3b). The experimentally obtained data is well described by Eq.(10) in the limit of $\zeta_m \ll 1$ if some additive term $\beta_0$ is introduced: $\beta = \frac{\omega_r}{\omega_0}\zeta_m + \beta_0$ (blue line in Fig. 3b). The presence of the term $\beta_0$ can be due to some inaquaracy in the determination of the incidence angle and due to the influence of the thermal effects which are not accounted in the theory above. It should be noted that the line slope is close to reciprocal of the refractive index of the magnetic film $1/n$ which agrees with Eq.(10). Indeed, for relatively large magnetic fields, where oscillation frequency is linear in $H$ the ratio of $\omega_r/\omega_0 \approx 1$ and $\beta \approx \zeta_m \approx \zeta_i/n$.

Change of the phase with variation of the incidence angle can be increased if $\omega_r$ is much larger than $\omega_0$. It might be possible for the films with large magnetic anisotropy. For example, for a film with $4\pi M = 1800$ Gs and $K_U = -1 \times 10^4$ erg/cm$^3$ in the external magnetic field $H = 10$ Oe, $\omega_r/\omega_0 = 13.9$ which implies $\beta = 78$ deg. for the incidence angle of $\zeta_i = 45$ deg.

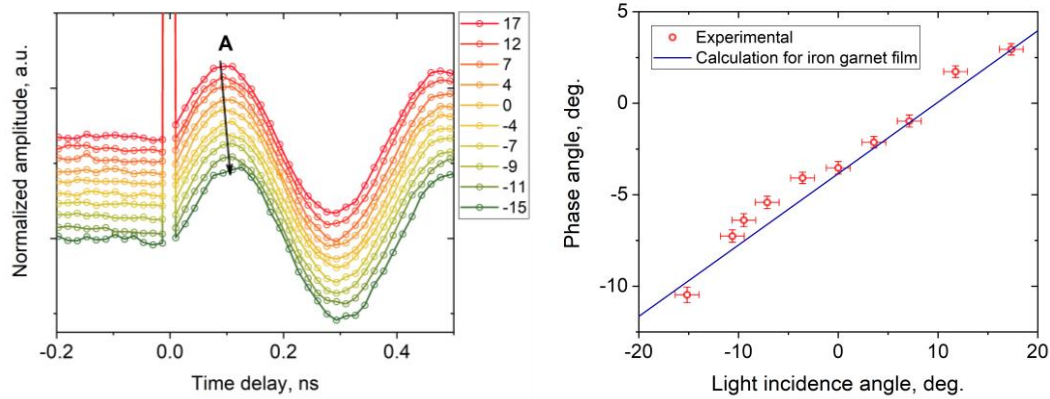

Fig. 3. a. Normalized time-resolved change of the Faraday rotation indicating the magnetization precession. The pump light incidence angle is varied from -15 to 17 degrees. The black line is the guide for the eye demonstrating the A peak position change. b. Dependence of the phase angle on the pump light incidence angle. External magnetic field is 850 Oe.

**Conclusion**

In conclusion, we have demonstrated the approach for the phase shift of the magnetostatic spin waves optically excited by circularly polarized laser pulses. It enables the precise control of the spin wave parameters for the case when the inverse Faraday effect is responsible for the origin of the ultrafast magnetization dynamics. The experimental data is in a good agreement with the proposed theoretical interpretation of the effect, rendering the linear trend for the phase angle change with the variation of the pump light incidence angle. We demonstrate the 15 deg. phase shift change with the variation of light incidence angle by 30 deg. The experimental sample film is not optimized for this effect. Calculations predict that the oscillation phase could be varied by almost 90 deg. in case when magnetic film of high in-plane magnetic anisotropy is placed in a relatively small external magnetic field of a few tens of Oe. This study provides a simple and effective way for the precise manipulation of ultrafast magnetization dynamics, which is of the increasing importance for spintronic and magnetophotonic applications.

**Funding**. Russian Science Foundation (17-72-20260).

spin wave generation in iron garnet by linearly polarized light pulses," J. Appl. Phys. **116**, (2014).